\documentclass{article}
\usepackage[utf8]{inputenc}
\usepackage{comment}
\usepackage{answers}
\usepackage{tikz}
\usepackage{ytableau}
\usepackage{caption}
\usepackage{setspace}
\usepackage{graphicx}
\usepackage{enumitem}
\usepackage{multicol}
\usepackage{amsfonts}
\usepackage{mathrsfs}
\usepackage{wrapfig}
\usepackage{multicol}
\usepackage[top = 2 cm, bottom = 2 cm, left = 1.5 cm, right = 1.5 cm]{geometry}
\usepackage{amsmath,amsthm,amssymb}

\usepackage[english]{babel}

\def\be{\begin{eqnarray}}
\def\ee{\end{eqnarray}}
\def\nn{\nonumber}
\def\p{\partial}
\def\tr{{\rm tr}\,}

\newcommand{\Tr}{\operatorname{Tr}}

\begin{document}

\title{\vspace{1.5cm}\bf
Phases in WLZZ matrix models}

\author{
A. Mironov$^{b,c,d,}$\footnote{mironov@lpi.ru,mironov@itep.ru},
A. Oreshina$^{a,c,}$\footnote{oreshina.aa@phystech.edu},
A. Popolitov$^{a,c,d,}$\footnote{popolit@gmail.com}
}

\date{ }

\maketitle

\vspace{-5.5cm}

\begin{center}
\hfill FIAN/TD-13/25\\
\hfill IITP/TH-21/25\\
\hfill ITEP/TH-24/25\\
\hfill MIPT/TH-17/25
\end{center}

\vspace{4.5cm}

\begin{center}
$^a$ {\small {\it MIPT, Dolgoprudny, 141701, Russia}}\\
$^b$ {\small {\it Lebedev Physics Institute, Moscow 119991, Russia}}\\
$^c$ {\small {\it NRC ``Kurchatov Institute", 123182, Moscow, Russia}}\\
$^d$ {\small {\it Institute for Information Transmission Problems, Moscow 127994, Russia}}
\end{center}

\vspace{.1cm}

\begin{abstract}
We discuss the space of solutions to the Ward identities associated with the WLZZ models. We mostly concentrate on the case of these models described by a two-matrix model with the cubic potential in one of the matrices. We study how this space of solutions can be described by the freedom in choosing integration contours in the matrix integral.
\end{abstract}

\section{Introduction}

One of the most informative object in string theory is the partition function, and one of the basic goals is studying its properties in various models. There are three possible definitions of the partition function: as a (functional) integral over possible trajectories, as a solution to an infinite set of non-linear differential equations (Ward identities), as a matrix element. An important example of the string theory partition function is provided by matrix model partition functions.

There are a few various possibilities of introducing matrix models, apart from giving an explicit matrix model integral. One of them is the $W$-representation\footnote{In fact, $W$-representations for generating partition functions is the standard idea, which goes back to \cite{DJKM,wrep,wrep1,wrep2}.} \cite{MSh,Wmore,Max,MMM}, which is, in fact, a direct consequence of Ward identities in the matrix model \cite{MMM}. In particular, a wide class of WLZZ matrix models \cite{China1,China2} can be realized by $W$-representations associated with commutative subalgebras of the $W_{1+\infty}$ algebra \cite{MMMP}. Moreover, their $\beta$-deformations \cite{China1,China2,Ch1,Ch2} and even $q,t$-deformations \cite{Ch3} can be realized by $W$-representations associated, respectively, with commutative subalgebras \cite{MMMP} of the algebra of affine Yangian \cite{Yang1,Yang2} and with those \cite{MMP} of the Ding-Iohara-Miki (DIM) algebra \cite{DI,Miki} (or of the elliptic Hall algebra \cite{K,BS,S}, which is basically the same \cite{S,Feigin}).

Note that the $W$-representation provides an unambiguous matrix model partition function given as a power series in their variables. For instance, one can define the Gaussian Hermitian one-matrix model as the matrix integral over $N\times N$ Hermitian matrix:
\be\label{ZG}
Z^{(GH)}(p_k;N)=\int dH \exp\left(-{\mu\over 2}\Tr H^2+\sum_k{p_k\over k}\Tr H^k\right)
\ee
which is understood as a formal power series in $p_k$'s, and the measure on the Hermitian matrices $dH$ is normalized so that $Z^{(GH)}(0;N)=1$.
At the same time, one can derive the Ward identities \cite{D,AMJ,MM,IM}
\be
\hat L_nZ^{(GH)}(p_k;N)&=&0,\ \ \ \ \ n\ge -1\nn\\
\hat L_n&:=&\sum_{k} (k+n)p_k\frac{\p}{\p p_{k+n}} + \sum_{a=1}^{n-1} a(n-a)\frac{\p^2}{\p p_a\p p_{n-a}}+\nn\\
&+& 2Nn\frac{\p}{\p p_n}    + N^2\delta_{n,0} + Np_1 \delta_{n+1,0}
- \mu (n+2)\frac{\p}{\p p_{n+2}}
\ee
and obtain from them that $Z^{(GH)}(p_k;N)$ satisfies the single equation \cite{MMM,Max}
\be\label{seq}
\sum_{n\ge 1}p_n\hat L_{n-2}Z^{(GH)}(p_k;N)&=&\left(2\hat W_{2}
- \mu\sum_{n\geq 1} np_n\frac{\p}{\p p_{n}} \right) Z\{p\} = 0\nn\\
\hat W_{2}:&=& {1\over 2} \sum_{k,n} (k+n)p_{n+2}p_k\frac{\p}{\p p_{k+n}} +{1\over 2} \sum_{a,b} ab p_{a+b+2}\frac{\p^2}{\p p_a\p p_b}+\nn\\
&+&N\sum_{k=3} (k-2)p_k \dfrac{\partial }{\partial p_{k-2}}
+ {N p_1^2\over 2}+{N^2 p_2\over 2}
\ee
This equation has the unique solution with $p_k$-series starting with 1, given by the $W$-representation for this matrix model partition function \cite{MMM,Max}
\be
Z^{(GH)}(p_k;N)=e^{{1\over\mu}\hat W_2}\cdot 1
\ee
which is equivalent to
\be\label{WG}
Z^{(GH)}(p_k;N)=\sum_R\mu^{-|R|/2}\xi_R(N)S_R\{N\}S_R\{\delta_{k,2}\}S_R\{p_k\}
\ee
where
\be\label{xi}
\xi_R(N):=\prod_{(i,j)\in R}(N+j-i)
\ee

However, in other cases, the Ward identities and the single equation have multiple solutions \cite{AMM1,AMM23,Max}. They can be obtained by changing the class of power series considered as solutions. Technically, in order to do this, one can shift, say, the variable $p_3\to p_3+g_3$ and allow $g_3$ to appear in denominators, i.e. one keeps $g_3$ an arbitrary complex parameter. Then, there are many solutions to the Ward identities, and only one of them survives in the limit $g_3\to 0$, which returns us back to $Z^{(GH)}(p_k;N)$.

The variety of solutions is ultimately related to different possibilities of choosing integration contours in the matrix integral \cite{ChM,CMMV,AMM1,AMM23,M1,M2}, which, in other terms, corresponds to choosing a phase of the model. In this paper, we study the space of solutions and the relation of the possible ambiguity of solutions to the Ward identities to the freedom of choosing integration contours in the case of WLZZ matrix models, which provide a more involved example being two-matrix models.

The paper is organized as follows. In section 2, we describe the basic facts on the variety of phases in the simplest example of the Hermitian one-matrix model. In section \ref{secWLZZ}, we remind the definition and the properties of the WLZZ models. Section 4 is devoted to describing solutions to the Ward identities in the case of cubic WLZZ model. An illustrative example the cubic model at $N=1$ is discussed in section \ref{secWLZZ32}. At last, section 6 contains some concluding remarks.

\paragraph*{Notation.}
We denote the Schur functions, which are graded polynomials of power sums $p_k:=\sum_{i=1}^Nx_i^k$, through $S_R\{p_k\}$. At the same time, if $x_i$ are eigenvalues of a matrix $X$, we use the notation $S_R(X)$. It can be also understood as a polynomial $S_R\{\tr X^k\}$. Here $R$ is the Young diagram (partition): $R_1\ge R_2\ge\ldots\ge R_{l_R}\ge 0$, and we sometimes use the term ``level" for size of the Young diagram $|R|=\sum_{i=1}^{l_R}$, which just gives the grading of the Schur function.

We also use the notation $S_{R/Q}\{p_k\}$ for the skew Schur functions defined from the relation
\be
S_R\{p_k+p'_k\}=\sum_QS_{R/Q}\{p_k\}S_Q\{p'_k\}
\ee
where $p_k$ and $p'_k$ are two sets of variables.

We also standardly define
\be
(n)!!:=\prod_{j=0}^{j< n/2} (n-2j)\nn\\
(n)!!!:=\prod_{j=0}^{j< n/3} (n-3j)
\ee
and
\be
\delta_{n|r}:={1\over r}\sum_{j=0}^{r-1}e^{2\pi inj\over r}
\ee
which means that $\delta_{n|r}$ is equal to 1 when $n$ is divisible by $r$ and is equal to zero otherwise.

The quantity $\xi_R(N)$ is defined in (\ref{xi}).

\section{The basic example: Hermitian one-matrix model}

We start our consideration with the basic example, which is well-studied and well-known: the Hermitian matrix model \cite{AMM1,AMM23,M1,M2}. At the level of matrix integral, the partition function of this model is formally given by the integral over the $N\times N$ Hermitian matrix $H$:
\be\label{pf}
Z^{(H)}(p_k;N)=\int dH \exp\left(\sum_k{p_k\over k}\Tr H^k\right)={1\over N!}\int\prod_{i=1}^Ndh_i\Delta(h)^2
\exp\left(\sum_{i,k}{p_k\over k}h_i^k\right)
\ee
where $\Delta(h):=\prod_{i<j}(h_i-h_j)$ is the Vandermonde determinant, $h_i$, $i=1,\ldots,N$ are the eigenvalues of the matrix $H$, and the second equality is a result of integration over the angular variables.

However, the integral (\ref{pf}) has yet to be defined, in particular, one has to specify the integration contour. If one integrates over all Hermitian matrices, the integral is generally not defined, and one has to say something about the parameters $p_k$. If, however, one somehow managed to define the integral, one can immediately derive an infinite set of Ward identities, which form a Borel subalgebra of the Virasoro algebra \cite{D,AMJ,MM,IM}:
\be\label{Vir}
\hat L_nZ^{(H)}(p_k;N)&=&0,\ \ \ \ \ n\ge -1\nn\\
\hat L_n&:=&\sum_{k} (k+n)p_k\frac{\p}{\p p_{k+n}} + \sum_{a=1}^{n-1} a(n-a)\frac{\p^2}{\p p_a\p p_{n-a}}+\nn\\
&+& 2Nn\frac{\p}{\p p_n}    + N^2\delta_{n,0} + Np_1 \delta_{n+1,0}
\ee
These Ward identities are derived in assumption that the integrand vanishes at the boundary.

Let us now turn the logic around: start from these Ward identities and solve them in order to see which matrix integrals one can obtain this way. To this end, first of all, one needs to say what kind of solutions of (\ref{Vir}) one is looking for. For instance, one may require that solution is a formal power series in $p_k$. Then, (\ref{Vir}) has the only solution $Z^{(H)}(p_k;N)=const$.

The next step is to assume that the solution is a power series in all $p_k$ but $p_2$. The simplest way to deal with this case is to shift $p_2\to g_2+p_2$ and to look for a solution which is a power series in all $p_k$ keeping $g_2$ finite. The solution in this case is unique:
\be
Z^{(H,2)}(p_k;N)=\sum_R(-g_2)^{-|R|/2}\xi_R(N)S_R\{N\}S_R\{\delta_{k,2}\}S_R\{p_k\}
\ee
and coincides \cite{MMsi,MMsum,MMrev} at negative $g_2=-\mu$ with the partition function of the Gaussian Hermitian matrix model (\ref{WG}). As follows from (\ref{pf}), this model is given by the matrix integral over all Hermitian matrices (\ref{ZG}):
\be
Z^{(H,2)}(p_k;N)=Z^{(GH)}(p_k;N)=\int dH \exp\left({g_2\over 2}\Tr H^2+\sum_k{p_k\over k}\Tr H^k\right)
\ee
and the integral this time is well-defined at negative $g_2$ provided the exponential of $p_k$'s is understood as expanded into the power series, i.e. one obtains a sum of the Gaussian moments.

Further extension is to switch on also a finite value of $g_3$ after the shift $p_3\to g_3+p_3$. In this case, the matrix model integral is
\be\label{pf3}
Z^{(H,3)}(p_k;N)=\int dH \exp\left({g_3\over 2}\Tr H^3+\sum_k{p_k\over k}\Tr H^k\right)
\ee
and the exponential of $p_k$'s is again understood as expanded into the power series. As for the integration contour, one can chose it, for instance, along the imaginary axis so that the matrix integral goes over the anti-Hermitian matrices.

There is, however, a possibility of choosing a more general contour, which is no longer associated with Hermitian or anti-Hermitian matrices, but (anti)Hermitian property is inessential for deriving the Ward identities and, hence, studying their solution does not require this. From now on, we are dealing with the integral (\ref{pf}) as just with an $N$-fold integral over variables $h_i$ with the integration contours chosen in such a way that the integral exists along with all moments, and such that the integrand vanishes at the ends of the contours. The integration contour in (\ref{pf3}) for any eigenvalue
can be chosen as a linear combination of two contours, e.g., associated with the Airy functions of the first and the second kinds \cite{Airy}. Moreover, the linear combinations can be chosen differently for different eigenvalues, which gives a freedom parameterized by $2^N$ parameters. These parameters just parameterize the space of solutions of the Ward identities (\ref{Vir}) in the case of the shifted variable $p_3$ \cite{ChM,CMMV,AMM1,AMM23,M1,M2,Max}. One can associate these different solutions with different phases of matrix model systems, and these phases are called Dijkgraaf-Vafa phases \cite{DV,Dav,Theisen}.

In the case of the shifted variable $p_m$, the contour can be chosen as a linear combination of $m-1$ basis contours. Moreover, one can also additionally shift all $p_k$ with $k<m$ without changing the contours. The detailed discussion of the relation between the integration contours and the number of solutions of the Ward identities can be found in \cite{ChM,CMMV,AMM1,AMM23,M1,M2,Max,Pop,MMMP,AMMbgw,Pop2}.
Below we repeat a similar analysis for the two matrix model associated with the WLZZ model. Note that the $\beta$-deformation of the model does not influence the analysis at all: though the Ward identities deform, neither the space of their solutions, nor the integration contours do.

\section{$WLZZ$ models}\label{secWLZZ}

In the Introduction, we described the logic when one starts from the matrix integral, then derives the Ward identities, the single equation and, finally, the $W$-representation. In the case of WLZZ models, the logic was inverse: the models were originally given by their $W$-representations in \cite{China1,China2}, and then they were realized by two-matrix integrals \cite{Ch1,Ch2}. These latter admit integrating out the angular variables and realization completely in terms of eigenvalues with the Harish-Chandra-Itzykson-Zuber integral involved. This allows one to immediately construct the $\beta$-deformation of the model \cite{MOP1,MOP2}. Here we use the approach of \cite{MOP1,MOP2}, derive the Ward identities, and then discuss solutions to these Ward identities after shifting variables.

\subsection{$W$-representation of WLZZ matrix models}

The WLZZ models were originally \cite{China1,China2}\footnote{In fact, in \cite{China1,China2} there were considered only the simplest WLZZ models associated with the commutative family of Hamiltonians of the rational Calogero system \cite{MMCal}. The whole class of WLZZ models associated with commutative subalgebras of the $W_{1+\infty}$ algebra \cite{MMMP} is described in \cite{Ch1,Ch2}. However, in this paper, we consider only the simplest models from \cite{China1,China2}.}  divided into three groups according to the grading of the operator giving the $W$-representation: positive, zero and negative. The corresponding partition functions are
        \begin{equation}
        \begin{cases} \label{WLZZPartFunc}
            Z^{(+)}(N;\bar p,p)= e^{\sum_k{1\over k}\bar p_k\hat{W}_k} \cdot 1, \ k>0 \\
            \\
            Z^{(0)}(N;p)= e^{\hat{W}_0} \cdot e^{\beta p_1}, \ m=0 \\
            \\
            Z^{(-)}(N;\bar p,p,g)= e^{\sum_k{1\over k}\bar p_k\hat{W}_{-k}} \cdot e^{\sum_k{g_kp_k\over k}}, \ k>0
        \end{cases}
        \end{equation}
where $\hat W_k$ are operators in variables $p_k$. In fact, they are infinite sums of the differential operators.

The manifest description of these operators is based on the cut-and-join operator $\hat W_0$ \cite{GJ,MMN}
        \begin{align}\label{W0b}
            \hat W_0 := \ \frac{1}{2}\sum_{a,b=1} \left(abp_{a+b}\frac{\partial^2}{\partial p_a \partial p_b} + (a+b) p_a p_b \frac{\partial}{\partial p_{a+b}}\right) +  N\sum_{a=1} ap_a\frac{\partial}{\partial p_a}+{ N^3\over 6}
            = {1\over 6}\sum_{a,b,c\in\mathbb{Z}}^{a+b+c=0}:p_ap_bp_c:
        \end{align}
where we denoted $p_0=N$,$\ p_{-k}=k{\partial \over\partial p_k}$, and the normal ordering means all the derivatives put to the right. The positive and negative $W$-operators can be constructed from the $W_0$  using
\be
\hat W_1 &= & \ [\hat W_0, p_1]\ =\ \frac{1}{2} \sum_{a+b=1} :p_{a} p_{b}:\nn\\
\hat W_{-1}&= & \ \left[{\p\over\p p_1},\hat W_0\right]\ =\ \sum_{b=0} p_b p_{-b-1}
\ee
and the generating operators
        \begin{align}\label{encrOp}
            \hat E = & \ [\hat W_0,\hat W_1] = \frac{1}{3} \sum_{a+b+c=1}^\infty :p_a p_b p_c:
             \\ \notag
            \hat{F} & = [\hat W_{-1},\hat W_0] = \frac{1}{3} \sum_{a+b+c = -1} : p_a p_b p_c:
            \label{lowOp}
        \end{align}
Then, all other $W$-operators are generated iteratively:
        \be\label{negativeBranchW}
            \hat W_{n+1}={1\over n}[\hat E,\hat W_n]\\
        \label{positiveBranchW}
            \hat W_{-n-1}={1\over n}[\hat W_{-n},\hat F]
        \ee

Direct calculation of the $W$-operators shows that $\hat W_2$ is exactly the $W$-operator of the Hermitian Gaussian model (\ref{WG}), $\hat W_0$ generates the Hurwitz partition function \cite{GJ,MMN}, and $\hat W_{-2}$, the Hermitian Gaussian model in the external field \cite{MSh}.

\subsection{Integral representation}

The integral representation for the WLZZ models can be presented in the form of the two-matrix model that interpolates between the positive and negative branches of the WLZZ models at once:
        \begin{equation}\label{WLZZfullInteralRep}
            Z(N;\bar p,p,g) =\int\int_{N\times N}[dXdY] \exp \left(-\Tr XY +\Tr Y\Lambda+ \sum_k {g_k\over k} \Tr X^k+\sum_k{\bar p_k\over k}\Tr Y^k\right)
        \end{equation}
where $p_k=\Tr\Lambda^k$. This partition function depends on three sets of variables $p_k$, $\bar p_k$ and $g_k$. As usual, the exponentials of these three sets of variables are understood as power series. The integral is normalized such that $Z(N;0,0,0)=1$. The two branches of the WLZZ models are now obtained from this matrix model as
        \begin{align}
            Z^{(+)}(N;\bar p,p)=Z(N;p,0,\bar p)\\
            Z^{(-)}(N;\bar p,p,g)=Z(N;\bar p,p,g) \notag
        \end{align}
i.e. the positive branch turns out to be a particular case of the negative branch.

This two-matrix integral \eqref{WLZZfullInteralRep} can be rewritten in the eigenvalue form \cite{MOP1,MOP2}:
        \begin{equation} \label{PartFuncEigen}
            Z(N;\bar p,p,g)=\int[dxdy]  \Delta^{2} (x) \Delta^{2} (y) I(X,-Y)  I(\Lambda,Y) \exp \left[ \sum_{k \geq 1} \frac{g_k}{k} \sum_{j=1}^{N} x_j^k +  \sum_{k \geq 1} \frac{\bar p_k}{k}\sum_{j=1}^{N} y_j^m \right]
        \end{equation}
where $I(X,Y)$ is the Harish-Chandra-Itzykson-Zuber integral \cite{HC,IZ}, which emerges after integrating out the angular variables and, in terms of eigenvalues, is equal to
        \begin{equation}\label{IZdet}
            I(X,Y) =\frac{\det_{j,k} \left[ e^{x_j y_k} \right]}{ \Delta(x) \Delta(y) }
        \end{equation}

Note that, in variance with the Hermitian one-matrix model case, the integral \eqref{PartFuncEigen} is {\bf well-defined without any shifts} of variables.
In \cite{MOP1}, we used as integration contours the real axis for the $x$-integration, and the imaginary axis for the $y$-integration. In fact, as it was demonstrated in \cite{MOP2}, the integral \eqref{PartFuncEigen} is almost {\bf independent of the integration contour}  (if exists at all): if all integration contours over $x_i$ and $y_i$ pass through the origin of coordinates, the integral does not depend on concrete choices of these contours, otherwise the integral vanishes. We additionally illustrate this claim in sec.5.2. The non-zero integral (i.e. for the contours passing through the origin) can be expressed in terms of Schur polynomials as
        \begin{equation}\label{SuperintegrabilityWLZZ}
            Z(N;\bar p,p,g)=\sum_{R,Q}{\xi_R(N)\over \xi_Q(N)}S_{R/Q}\{\bar p_k\}S_R\{g_k\}S_Q\{p_k\}
        \end{equation}
        This form of the partition function is associated with the superintegrability property \cite{MMrev,MMZ}.

Note that, since this unshifted case provides (in fact, only in this case) the unique solution of the Ward identities (see the next section), one may think it is a kind of counterpart of the Gaussian Hermitian one-matrix model, when the variable $p_2$ is shifted.

We consider below only the case cubic in $X$ in order to make all calculations more explicit and observable. Moreover, for the sake of simplicity, we deal with the positive branch of the WLZZ models\footnote{The two-matrix model for the positive branch was actually first introduced in \cite{AMMN}.}. Nevertheless, our results can be easily extended to the negative branch by substituting the polynomial form of the Harish-Chandra-Itzykson-Zuber integral \cite{Kazak,Bal,Mor}
        \begin{equation}\label{IZSchur}
            I(\Lambda,Y) = \sum_{R} {1\over\xi_R(N)}S_R(\Lambda) S_R(Y)
        \end{equation}

\subsection{Ward identities}\label{secWI}

Here we explore the space of Ward identity solutions, following the general scheme. The technique of Ward identities for the positive branch two-matrix model was developed in \cite{Gava}. The Ward identities are looking as follows:
    \begin{equation}\label{nb}
        (n+1){\partial Z^{(+)}(N;\delta_{k,m},p)\over\partial p_{n+1}}=\hat{\widetilde W}^{(m)}_{n-m+1} Z^{(+)}(N;\delta_{k,m},p)
    \end{equation}
where $\hat{\widetilde W}^{(m)}_{n-m+1}$ are generators of the $\widetilde W$-algebra \cite{Gava,GKMU,DMP,Dr}. There is no general formula for these generators, however, they can be calculated at every particular $m$. Now we deal with the $m=3$ case.

The Ward identities for each $k\geq 0$ in the cubic case are \cite{Gava}
        \begin{equation}\label{nb3}
            (n+1){\partial Z^{(+)}(N;\delta_{k,3},p)\over\partial p_{n+1}}=\hat{\widetilde W}^{(3)}_{n-2} Z^{(+)}(N;\delta_{k,3},p)
        \end{equation}
and
        \begin{equation}\label{W3b}\begin{array}{c}
            \hat{\widetilde W}^{(3)}_{n}= \frac{ n(n-2)}{ 2} N \delta_{n,2}+ N p_1^2\delta_{n,0}+\frac{ n(n-1)(n-2)}{2} \frac{\partial}{\partial p_{n-2}} +\cr
            \cr
            2\sum_{k}p_k \sum_{a=0}^{p+k-2} a(n+k-2-a) \frac{\partial^2}{\partial p_{a} \partial p_{n+k-2-a}}-\sum_{k}p_k \sum_{a=0}^{k-2} a(n+k-2-a) \frac{\partial^2}{\partial p_{a} \partial p_{n+k-2-a}} +\cr
            \cr
            + \sum_{a=0}^{n-2} \left[ \frac{1}{2}n - (a+1) \right] \frac{\partial^2}{\partial p_{a} \partial p_{n-2-a}}+ \sum_{k,l} p_k p_l \frac{\partial}{\partial p_{n+k+l-2}}+ \sum_{a+b+c=n-2} abc \frac{\partial^3}{\partial p_{a}\partial p_{b}\partial p_{c}}
        \end{array}
        \end{equation}

\section{The space of solutions of the cubic WLZZ model}

\subsection{The simplest case: a unique solution to the Ward identities}

Let us look for a solution $Z_3(N;p)$ to \eqref{nb3} which is a graded power series in $p_k$, i.e. it is generally of the form
        \begin{equation}\label{ansatz}
            Z_3(N;p)=\sum_{R} b_R S_R\{p_k\}
        \end{equation}
with some coefficients $b_R$, which can be evaluated from the Ward identities iteratively level by level. They are determined from the Ward identities uniquely and are exactly equal to the superintegrability coefficients in \eqref{SuperintegrabilityWLZZ} at $p_k=0$, $g_k=\delta_{k,3}$, $\bar p_k=p_k$, i.e. \eqref{SuperintegrabilityWLZZ} reduces to
\be\label{si3}
Z_3(N;p)=\sum_{R,Q}\xi_R(N)S_{R}\{\delta_{k,3}\}S_R\{p_k\}
\ee
A few first coefficients are
        \be\label{bRWLZZ3}
            &\phantom{.}&b_{\varnothing}=1, \nn\\
            \hline\hline\nn\\
            &\phantom{.}&b_{[3]}={N (N+1) (N+2)\over 3}, \ b_{[2,1]}=-{(N-1) N (N+1)\over 3},
            \ b_{[1,1,1]}={(-2 + N) (-1 + N) N\over 3},\nn\\
            \hline\hline\nn\\
            &\phantom{.}&b_{[6]}={N (1 + N) (2 + N) (3 + N) (4 + N) (5 + N)\over 18}, \ b_{[5,1]}=-{(-1 + N) N (1 + N) (2 + N) (3 + N) (4 + N)\over 18},\nn\\
            &\phantom{.}&b_{[4,1,1]}={(-2 + N) (-1 + N) N (1 + N) (2 + N) (3 + N)\over 18}, \ b_{[3,3]}={(-1 + N) N^2 (1 + N)^2 (2 + N)\over 9},\nn\\
            &\phantom{.}&b_{[3,2,1]}=-{(-2 + N) (-1 + N) N^2 (1 + N) (2 + N)\over 9}, \ b_{[3,1,1,1]}={(-3 + N) (-2 + N) (-1 + N) N (1 + N) (2 + N)\over 18},\nn\\
            &\phantom{.}&b_{[2,2,2]}={(-2 + N) (-1 + N)^2 N^2 (1 + N)\over 9}, b_{[2,1,1,1,1]}=-{(-4 + N) (-3 + N) (-2 + N) (-1 + N) N (1 + N)\over 18},\nn\\
            &\phantom{.}&b_{[1,1,1,1,1,1]}={(-5 + N) (-4 + N) (-3 + N) (-2 + N) (-1 + N) N\over 18},\nn\\
            \hline\hline\nn\\
            &\phantom{.}&\ldots
        \ee
All remaining coefficients at the first six levels are zero. The fact that the coefficients $b_R$ are unambiguously determined
means that the Ward identities have the unique solution. This solution is much similar to that in the Hermitian one-matrix model in the Gaussian case. Now we are going to make shifts of variables.

        \subsection{Solutions in the case of shifted variables}

As we discussed in sec.2, in order to generate a non-trivial space of solutions, one can shift variables: $p_k \to p_k+\alpha^m \delta_{m,k}$ with an arbitrary (graded) parameter $\alpha$. The integral representation of the cubic model in this case is
        \begin{equation}\label{Z32}
            Z_{3,m}(N;p) = \int\int_{N\times N}[dXdY]\exp\left(-\Tr XY+ {\alpha^m \over m}\Tr Y^m+ {1\over 3}\Tr X^3+\sum_k{ p_k\over k}\Tr Y^k\right)
        \end{equation}
and the Ward identities are changed respectively so that the $\tilde W$-algebra generators become
        \begin{equation}\label{Virasoro3m}
            \begin{array}{c}\hat{\widetilde W}^{(3,m)}_{n}= n(n-1)(n-2) \frac{1}{2} \frac{\partial}{\partial p_{n-2}} + n(n-2) \frac{ N }{ 2} \delta_{n,2} + N p_1^2 \delta_{n,0}+ \sum_{a+b+c=n-2} abc \frac{\partial^3}{\partial p_{a}\partial p_{b}\partial p_{c}} +\cr
            \cr
            +2\sum_{k} p_k \sum_{a=0}^{n+k-2} a(n+k-2-a) \frac{\partial^2}{\partial p_{a} \partial p_{n+k-2- a}} +  \underline{2 \alpha^m \sum_{a=0}^{n+m-2} a(n+m-2-a) \frac{\partial^2}{\partial p_{a} \partial p_{n+m-2-a}}}-\cr
            \cr
            -\sum_{k}p_k \sum_{a=0}^{k-2} a(n+k-2-a) \frac{\partial^2}{\partial p_{a} \partial p_{n+k-2-a}} -  \underline{ \alpha^m  \sum_{a=0}^{m-2} a(n+m-2-a) \frac{\partial^2}{\partial p_{a} \partial p_{n+m-2-a}}} +\cr
            \cr
            + \sum_{a=0}^{n-2} \left[ \frac{1}{2}n - (a+1) \right] \frac{\partial^2}{\partial p_{a} \partial p_{n-2-a}}+ \sum_{k,l}p_k p_l \frac{\partial}{\partial p_{n+k+l-2}}+  \underline{ 2\alpha^m \sum_{k}p_k\frac{\partial}{\partial p_{n+k+m-2}}+ \alpha^{2m} \frac{\partial}{\partial p_{n+2m-2}}}
            \end{array}
        \end{equation}
Underlined are the terms emerging after the shift. Summing up the Ward identities \eqref{nb3} weighted with $p_{n+1}$, one obtains the single equation, which is a counterpart of (\ref{seq}):
        \begin{equation}\label{SigleEq}
            \left(\sum_{n=0}^{\infty}(n+1)p_{n+1}{\partial \over\partial p_{n+1}} - \sum_{n=0}^{\infty} p_{n+1}\hat{\widetilde W}^{(3,m)}_{n-2} \right)Z_{3,m} = 0
        \end{equation}
Substituting the shifted generators $\hat{\widetilde W}^{(3,m)}_{n-2} $, one observes that the single equation contains differential operators of four different gradings:
        \begin{equation}\label{SE3}
            \left(\sum_{n=0}^{+\infty} (n+1) p_{n+1} {\partial \over \partial p_{n+1}}- \sum_{n=2}^{+\infty} p_{n+1} \hat{\widetilde W}^{(3)}_{n-2}\ \underline{-\hat O_{3-m}-\hat O_{3-2m} } \right)Z_{3,m}=0
        \end{equation}
where the terms $\hat O_{3-m}$ and $\hat O_{3-2m}$ have gradings $3-m$ and $3-2m$ respectively, i.e. contains distinct powers of $\alpha$:
        \begin{multline}
            \hat O_{3-m}=  \alpha^m \sum_{n=0}^{+\infty} p_{n+1} \left[2 \sum_{k}p_k\frac{\partial}{\partial p_{n+k+m-2}}- \sum_{a=1}^{m-2} a(n+m-2-a) \frac{\partial^2}{\partial p_{a} \partial p_{n+m-2-a}}\right. +\\
            +\left. N (n+m-2) \frac{\partial}{ \partial p_{n+m-2}}+2  \sum_{a=1}^{n+m-2} a(n+m-2-a) \frac{\partial^2}{\partial p_{a} \partial p_{n+m-2-a}}  \right]
        \end{multline}
        \begin{equation}
             \hat O_{3-2m}=\alpha^{2m} \sum_{n=0}^{+\infty} p_{n+1}   \frac{\partial}{\partial p_{n+2m-2}}
        \end{equation}
The existence of these operators is the main reason of non-uniqueness of the Ward identity solution: substituting the general power series
        \begin{equation} \label{ansatzAlpha}
            Z_{3,m}=\sum_{R}b_{R}(\alpha) S_{R}\{p\}
        \end{equation}
into the Ward identities (\ref{SE3}), and solving them iteratively, one realizes that, at every level, the number of coefficients $b_R$ is bigger than the number of equations. In fact, the number of equations depends of the term with minimal grading $\hat O_{3-2m}$. It can be easily checked that the number of equations at level $|R|$ is equal to $|R|+(3-2m)$, which is much less than $p(|R|)$, the number of partitions of the level $|R|$, which gives the number of distinct coefficients $b_R$.

\subsection{The freedom in solutions}

Let us illustrate the freedom in the coefficients $b_R$, i.e. describe the space of solutions in the simplest case of $Z_{3,2}$ model.

When solving the Ward identities, we insert the power series \eqref{ansatzAlpha} into the \eqref{SE32} and successively put to zero the coefficients in front of the monomials $p_{\Delta}= p_1^{\Delta_1} ...p_k^{\Delta_k}...$. The coefficient of $p_\emptyset$ is identically zero. The coefficient at level 1, i.e. in front of $p_1$, gives rise to the only equation
    \begin{equation}\label{l1}
        \alpha ^4 b_{[1,1]} -\alpha ^4 b_{[2]}+b_{[1]}-N^2\alpha^2b_{\emptyset}=0
    \end{equation}
We consider it as an equation for $b_R$ with $|R|=2$. Thus, one of the two coefficients at level $|R|=2$ remains arbitrary. Note that $b_\emptyset$ is fixed by the normalization of the partition function, which we choose, without loss of generality, to be 1 in the case of matrix model integrals.

Similarly, at level 2, there are two equations (coefficients in front of $p_2$ and $p_1^2$) for $b_R$ with $|R|=3$:
        \begin{equation}\label{l2}
            \begin{cases}
                \alpha ^4 b_{[2,1]}-2\alpha ^4 b_{[1,1,1]}-2b_{[1,1]}-2\alpha ^4 b_{[3]}-(2+3N+N^2) \alpha ^2 b_{[1]}+2b_{[2]}=0
                \\
                \alpha ^4 b_{[1,1,1]}+b_{[1,1]}-\alpha ^4 b_{[3]}-(2+N^2) \alpha ^2 b_{[1]}+b_{[2]}=0
            \end{cases}
        \end{equation}
Generally, at level $k$, one obtains $p(k)$ equations for the coefficients $b_R$ with $|R|=k+1$. Here $p(k)$ is the number of partitions of integer $k$ (we put $p(0):=0$).

In other words, one observes that the number of constraints for the coefficients $b_R$ with $|R|\le K$ (there are totally $\sum_{k=1}^{K}p(k)$ these coefficients) is equal to $\sum_{k=1}^{K-1}p(k)$, and the number of the coefficients remaining unconstrained is equal to
        \begin{equation}\label{N1}
            \mathbf{N}_{K} = \sum_{k=1}^{K}p(k)-\sum_{k=1}^{K-1}p(k)=p(K)
        \end{equation}
Thus, there are a lot of free parameters in solution to the Ward identities, the space of solutions is huge and can not be associated with the freedom in choosing integration contours, since their number is finite at finite $N$.

However, formula (\ref{N1}) along with formula (\ref{ansatzAlpha}) can be associated with the matrix model partition function only in the case of large enough $N$, say, $N>K$. Equivalently, one can consider the limit of $N\to \infty$. The reason is that only the first $N$ traces of powers of the $N\times N$ matrix are independent, which means that only first $N$ variables $p_k$ in (\ref{WLZZfullInteralRep}) are ``truly" independent variables in the sense of D-module, i.e. all the derivatives in these variables are independent. All other variables are involved in additional constraints for the partition function. For instance, in the case of $N=1$, these additional constraints are:
        \begin{equation}
            \frac{\partial Z}{\partial p_n}= \frac{\partial^n Z}{\partial p_1^n}
        \end{equation}
since
        \begin{equation}
            Z \sim \int dx \exp\left[p_1 x_1+ ... +p_n x_1^n+... \right]
        \end{equation}
In the case of $N=2$, these constraints are more involved: the first of them is
        \begin{equation}
            \left(\frac{\partial^2 }{\partial p_1\partial p_2}-2 \frac{\partial }{\partial p_3}\right) Z= \frac{\partial^3 Z}{\partial p_1^3}
        \end{equation}
These additional constraints are described by the condition
        \begin{equation}
            S_{R}\Big(Y_{N\times N}\Big)=0
        \end{equation}
if $l_R>N$. In other words, in the matrix model partition function, there is an additional restriction in the sum in (\ref{ansatzAlpha}): it goes over all partitions $R$ with $l_R\le N$. Then, the formula for $\mathbf{N}_{K}$ has to be changed accordingly:
        \begin{equation}\label{NumberOfEq}
            \mathbf{N}_{K,N} = p(K)-\sum_{k=1}^K\Big(p(k)-p(k,N)\Big)
        \end{equation}
where $p(k,N)$ is number of partitions of $k$ with number\footnote{The quantity $p(k,N)$ can be described by the generating function
       \begin{equation}
           g(z,N):=\sum_{k}p(k,N)z^k=\prod_{i=1}^{N}\frac{1}{1-z^i}
        \end{equation}
which reduces to the generation function of $p(k)$ at $N\to\infty$:
        \begin{equation}
        g(z):=\sum_{k}p(k)z^k= \prod_{i=1}^{\infty}\frac{1}{1-z^i}
        \end{equation}
In the case of $N=1$, $p(k,1)=1$ independently of $k$.} of parts no more than $N$, i.e. with $l_R\le N$.

The key point here is that, at some $K_0$, the quantity $\mathbf{N}_{K_0,N}$ becomes negative, and the system of equations becomes over-determined. It is still solvable, since some of equations are no longer independent.

In order to see how this works, consider, for instance, the examples of $N=1$. Then, equation (\ref{l1}) determines $b_{[2]}$, and the two equations (\ref{l2}) are equivalent and determine $b_{[3]}$, etc. Hence, totally, there is only one free parameter $b_{[1]}$ in this case. As we will explain in the next section, this parameter corresponds exactly to the freedom in choosing the integration contours of one-dimensional integral.

Similarly, in the example of $N=2$, one determines from (\ref{l1}) the coefficient $b_{[2]}$, while $b_{[1]}$ and $b_{[1,1]}$ remain free parameters, and (\ref{l2}) determine $b_{[3]}$ and $b_{[2,1]}$, and all other equations do not admit more free parameters. Hence, there are exactly two free parameters in this case associated with two arbitrary integration contours. And, generally, there are $N$ arbitrary parameters associated with $N$ arbitrary integration contours.

Hence, our claim is that {\bf only in the limit of $N\to\infty$ the space of solutions to the Ward identities is described by a freedom in choosing the integration contours in the matrix model partition function}.

The next section illustrates our description of the $Z_{3,m}$ models in the simplest nontrivial example of $Z_{3,2}$ model with $N=1$, where the story with the contour becomes completely explicit.

\section{$Z_{3,2}$ at $N=1$\label{secWLZZ32}}

In this section, we study in detail the cubic model with the second variable $p_2$ shifted, i.e. with the Gaussian potential in $Y$ in the one-dimensional ($N=1$) case. The matrix model representation of the partition function in this case is an ordinary integral
    \begin{equation}\label{Z32N1}
        Z_{3,2}(p) = \int\int dx dy \exp\left(- x y+ {\alpha^2 y^2 \over 2} + {x^3\over 3} +\sum_k { p_k\over k} y^k\right)
    \end{equation}
This partition function is generally singular at $\alpha \to 0 $ (i.e. the integral does not exist), and there is a unique regular solution, which reduces to $Z_3(N;p)$ at this limit (in accordance with the claim in sec.3.2).

First, let us look at the superintegrability formula (\ref{si3}). As soon as we deal with the $N=1$ case, only the one-line Young diagrams contribute ($\xi_R(1)=0$ for multi-line Young diagrams), and one obtains
    \be\label{Z32Schur}
        Z_{3}(1;p)&=&\sum_{n}\xi_{[n]}(1) S_{[n]} \{ \delta_{k,3} \} S_{[n]} \{ p_k+\alpha^2 \delta_{2,k}\}=
        \sum_{n\ge r} \alpha^{n-r} \xi_{[n]}(1) S_{[n]}\{\delta_{k,3}\} S_{[n]/[r]} \{\delta_{2,k} \} S_{[r]}\{p_k\}=\nn\\
        &=&\sum_{r} S_{[r]} (p_k) \sum_{n\ge r} \alpha^{n-r} \frac{n!}{\left( n-r \right)! !\left( n\right)!!!} \delta_{n+r|2} \delta_{n|3}
    \ee
where we used that only the skew Schur functions $S_{[n]/Q}$ with one-line Young diagrams $Q$ are non-vanishing.

Now we discuss solutions to the Ward identities, and compare them with those obtained from the integral (\ref{Z32N1}) with different choices of the integration contours.

    \subsection{Ward identities solution}

According to \eqref{Virasoro3m}, the $\widetilde W$-generators are
    \begin{equation}
        \begin{array}{c}\hat{\widetilde W}^{(3,2)}_{n-2}=n(n-1)(n-2) \frac{ 1}{ 2 } \frac{\partial}{\partial p_{n-2}} + \frac{n(n-2)}{ 2} \delta_{n,2}+ p_1^2\delta_{n,0}+\cr
        \cr
        +2\sum_{k}( p_k+\alpha^2 \delta_{2,k}) \sum_{a=0}^{p+k-2} a(n+k-2-a) \frac{\partial^2}{\partial p_{a} \partial p_{n+k-2-a}}-\sum_{k}( p_k+\alpha^2 \delta_{2,k}) \sum_{a=0}^{k-2} a(n+k-2-a) \frac{\partial^2}{\partial p_{a} \partial p_{n+k-2-a}} +\cr
        \cr
        + \sum_{a=0}^{n-2} \left[ \frac{1}{2}n - (a+1) \right] \frac{\partial^2}{\partial p_{a} \partial p_{n-2-a}}+\sum_{k,l}( p_k+\alpha^2 \delta_{2,k})( p_l+\alpha^2 \delta_{2,l}) \frac{\partial}{\partial p_{n+k+l-2}}+ \sum_{a+b+c=n-2} abc \frac{\partial^3}{\partial p_{a}\partial p_{b}\partial p_{c}} =\cr
        \cr
        =\frac{n(n-1)(n-2)}{ 2 } \frac{\partial}{\partial p_{n-2}} + \frac{n(n-2) }{ 2} \delta_{n,2}+ p_1^2 \delta_{n,0}+\cr
        \cr
        +2\sum_{k} p_k \sum_{a=0}^{p+k-2} a(n+k-2-a) \frac{\partial^2}{\partial p_{a} \partial p_{n+k-2-a}}- \sum_{k} p_k \sum_{a=0}^{k-2} a(n+k-2-a) \frac{\partial^2}{\partial p_{a} \partial p_{n+k-2-a}} +\cr
        \cr
        + \sum_{a=0}^{n-2} \left[ \frac{1}{2}n - (a+1) \right] \frac{\partial^2}{\partial p_{a} \partial p_{n-2-a}}+ \sum_{k,l} p_k p_l \frac{\partial}{\partial p_{n+k+l-2}}+ \sum_{a+b+c=n-2} abc \frac{\partial^3}{\partial p_{a}\partial p_{b}\partial p_{c}}+\cr
        \cr
        +\underline{ 2 \alpha^2 \sum_{a=0}^{p} a(n-a) \frac{\partial^2}{\partial p_{a} \partial p_{n-a}}- \alpha^2 \ n \frac{\partial}{ \partial p_{n}}+ \alpha^2 \sum_{k} p_k\frac{\partial}{\partial p_{n+k}}+ \alpha^4  \frac{\partial}{\partial p_{n+2}}}
        \end{array}
    \end{equation}
The corresponding single equation, obtained after summation of the Ward identities is
    \begin{equation}\label{SE32}
        \left(\sum_{n=0}^{+\infty} (n+1) p_{n+1} {\partial \over \partial p_{n+1}}- \sum_{n=2}^{+\infty} p_{n+1} \hat{\widetilde W}^{(3)}_{n-2} \ \underline{-\alpha^2 \hat O_1- \alpha^4 \hat O_{-1}} \right)Z_{3+2}=0
    \end{equation}
Here $\hat O_1$ and $\hat O_{-1}$ have gradings $1$ and $-1$ respectively, and the powers of $\alpha$ in front of them are, accordingly, distinct:
    \begin{equation}
         \hat O_1=  \sum_{n=2}^{+\infty} p_{n+1} \left[2\sum_{a=1}^{n-1} a(n-a) \frac{\partial^2}{\partial p_{a} \partial p_{n-a}} +3 n \ N \frac{\partial}{\partial p_n} + 2\sum_{k} p_k \frac{\partial}{\partial p_{n+k}}\right]
    \end{equation}
    \begin{equation}
         \hat O_{-1}= \sum_{n=2}^{+\infty} p_{n+1}  \frac{\partial}{\partial p_{n+2}}
    \end{equation}
Solving iteratively \eqref{SE32} with expansion \eqref{ansatzAlpha}, one obtains the coefficients $b_{R}(\alpha)$. A few first of them are
        \begin{multline}\label{bRWLZZ32}
            \ b_{[1,1]} = -\frac{b_{[1]}}{\alpha ^4} + b_{[2]}+ \frac{b_\emptyset}{\alpha ^2}, \ b_{[2,1]} = \frac{6 b_{[1]}}{\alpha ^2}- \frac{2 b_{[2]}}{\alpha ^4} + 2 b_{[3]}, \ b_{[1,1,1]} = \frac{3 b_{[1]}}{\alpha^2} +\frac{b_{[1]}}{\alpha ^8}- \frac{2 b_{[2]}}{\alpha ^4}+ b_{[3]} -\frac{b_\emptyset}{\alpha ^6},\\
            b_{[3,1]} = \frac{15 b_{[2]}}{\alpha ^2}- \frac{3 b_{[3]}}{\alpha ^4}+3 b_{[4]} +\frac{6b_\emptyset}{\alpha ^4}, \ b_{[2,1,1]} = -\frac{12 b_{[1]}}{\alpha ^6} +\frac{18 b_{[2]}}{\alpha ^2} +\frac{3 b_{[2]}}{\alpha ^8} -\frac{6 b_{[3]}}{\alpha ^4} +3 b_{[4]}+ \frac{9b_\emptyset}{\alpha ^4},\\
            b_{[1,1,1,1]} = -\frac{8 b_{[1]}}{\alpha ^6}- \frac{b_{[1]}}{\alpha ^{12}}+ \frac{7 b_{[2]}}{\alpha ^2}+ \frac{3 b_{[2]}}{\alpha ^8}- \frac{3 b_{[3]}}{\alpha ^4}+ b_{[4]} +\frac{4b_\emptyset}{\alpha ^4}+ \frac{b_\emptyset}{\alpha ^{10}},\
            \ldots
        \end{multline}
These equations are homogeneous, which reflects the fact the normalization of the partition function is a free parameter.

Though the superintegrability solution (\ref{Z32Schur}) at $N=1$ is associated only with the one-line Young diagrams, the general solution to the Ward identities is not: arbitrary Schur polynomials contribute to the sum (\ref{ansatzAlpha}). However, as we explained in s.4.3, these solutions are not associated with a $1\times 1$ matrix integral. Those associated with the matrix integral for $Z_3(N;p)$ (\ref{Z32Schur}) require $b_R=0$ if $l_R>1$, and
\be\label{bRWLZZ32mi}
\!\!\!\!\!\!\!\!\!\!\!\!\!\!\!\!&\phantom{.}&b_{[2]}=\frac{b_{[1]}}{\alpha ^4} - \frac{b_\emptyset}{\alpha ^2}, \ \ \ \ \ b_{[3]} =\left({1\over\alpha^8}-{3\over\alpha^2}\right)b_{[1]}
-{b_\emptyset\over\alpha^6},\ \ \ \ \ b_{[4]} = \left({1\over\alpha^{12}}-{8\over\alpha^6}\right)b_{[1]}-\left({1\over\alpha^{10}}-{3\over\alpha^4}\right)b_\emptyset,\\
\!\!\!\!\!\!\!\!\!\!\!\!\!\!\!\!&\phantom{.}&b_{[5]}= \left({1\over\alpha^{16}}-{15\over\alpha^{10}}+{15\over\alpha^4}\right)b_{[1]}-\left(          {1\over\alpha^{14}}-{10\over\alpha^8}\right)b_\emptyset,\ \ \ \ \
            b_{[6]}= \left({1\over\alpha^{20}}-{24\over\alpha^{14}}+{75\over\alpha^8}\right)b_{[1]}-
\left({1\over \alpha^{18}}-{19\over\alpha^{12}}+{15\over\alpha^6}\right)b_{[0]},
\ \ \ \ \ \ldots\nn
        \ee
One can easily check that a particular solution of these equations is given by
\be\label{an}
b_{[r]}=N\cdot\sum_{n\ge r} \alpha^{n-r} \frac{n!}{\left( n-r \right)! !\left( n\right)!!!} \delta_{n+r|2} \delta_{n|3}
\ee
where $N$ is an arbitrary normalization factor. For instance:
\be
\frac{b_{[1]}}{\alpha ^4} - \frac{b_\emptyset}{\alpha ^2}=N\cdot
\left(\sum_{k\ge 0} \alpha^{6k-2} \frac{(6k+3)!}{\left( 6k+2 \right)! !\left( 6k+3\right)!!!}-
\sum_{k\ge 0} \alpha^{6k-2} \frac{6k!}{\left( 6k \right)! !\left( 6k\right)!!!}\right)=\nn\\
=N\cdot\sum_{k\ge 0} \alpha^{6k-2} 6k\frac{6k!}{\left( 6k \right)! !\left( 6k\right)!!!}=
N\cdot\sum_{k\ge 0} \alpha^{6k+4} \frac{(6k+6)!}{\left( 6k+4 \right)! !\left( 6k+6\right)!!!}=b_{[2]}
\ee
etc.

The answer (\ref{an}) coincides with the coefficients from (\ref{Z32Schur}) upon choosing the unit normalization parameter $N=1$. In the next subsection, we obtain a general solution in this matrix integral case.

\subsection{Integration contours}

Now we are going to evaluate the ``matrix" integral for the model $Z_{3,2}$ in the $N=1$ case, when it is a one-dimensional integral
        \begin{equation}\label{z23n1}
            Z_{3,2}(p) = \int\int dx \ dy \exp\left(-xy+ {\alpha^2 \over 2} y^2+ {1\over 3} x^3+\sum_k{ p_k\over k} y^k \right)
        \end{equation}
This integral can be rewritten in the form
\begin{multline}\label{Z32F}
            Z_{3,2}(p)
            =\sum_{r} S_{[r]}(p_k)\int\int dx \ dy \exp\left(-xy+ {\alpha^2 \over 2} y^2+ {1\over 3} x^3\right) \ y^r=\\
            =\sum_{r} S_{[r]}(p_k)\left. \left[ \frac{\partial^r}{\partial p_1^r}
            \underbrace{\int\int dx \ dy \exp\left(-xy+ {\alpha^2 \over 2} y^2+ {1\over 3} x^3 + p_1 y\right)}_{F(p_1)} \right]  \right|_{p_1=0}
        \end{multline}
In order to compare it with formulas like (\ref{Z32Schur}), we need to construct an expansion of this integral in positive powers of $\alpha$. To this end, we rescale the integration variables $(x,y)\to(\frac{x}{\alpha^2},\frac{y}{\alpha^4})$ so that
        \begin{multline}
            F(p_1) = \int\int dx \ dy \exp\left(-xy+ {\alpha^2 \over 2} y^2+ {1\over 3} x^3+p_1 y \right) = \\
            =\frac{1}{\alpha^{6}}\int\int dx \ dy \exp\left(\frac{1}{\alpha^{6}}(-xy+ {y^2 \over 2} + {1\over 3} x^3 +\alpha^2 p_1 y)\right)
        \end{multline}
Now one can use the Laplace's method \cite{Lap}, and evaluate $F(p_1)$ in the neighbourhoods of the two saddle points of the integrand exponential:
        \begin{center}
            \begin{tabular}{c c}
                $x_1=\frac{1-\sqrt{1-4p_1\alpha^2}}{2}$ & $y_1=x_1^2$ \\
                $x_2=\frac{1+\sqrt{1-4p_1\alpha^2}}{2}$ & $y_2=x_2^2$
            \end{tabular}
        \end{center}
The integral is a sum of contributions of the two contours associated with these saddle points:
        \begin{multline}\label{sum}
            F(p_1) = \beta_1F^{(x_1,y_1)}(p_1) +\beta_2F^{(x_2,y_2)}(p_1)=\\
            =\frac{\beta_1}{\alpha^{6}}\int_{C_1} dx \ dy \exp\left(\frac{1}{\alpha^{6}}(-xy+ {y^2 \over 2} + {1\over 3} x^3 +\alpha^2 p_1 y)\right) +\frac{\beta_2}{\alpha^{6}}\int_{C_2} dx \ dy \exp\left(\frac{1}{\alpha^{6}}(-xy+ {y^2 \over 2} + {1\over 3} x^3 +\alpha^2 p_1 y)\right)
        \end{multline}
where $\beta_{1,2}$ are arbitrary constants regulating contributions of each integration contour.

The contribution associated with the first saddle point is
        \begin{equation}\label{F1}
            F^{(x_1,y_1)}(p_1) = e^{f(x_1,y_1)}\sum_{n=0} \alpha^{6n} \frac{(6n+1)!! }{(6n)!!!(6n+1)}\left(1-4p_1\alpha^2 \right)^{-\frac{6n+1}{4}}
        \end{equation}
where
        \begin{equation}
            f(x,y)=-xy+ {y^2 \over 2} + {1\over 3} x^3 +\alpha^2 p_1 y
        \end{equation}
This generating function after substituting to the \eqref{Z32F} produces the series that coincides with the superintegrability solution \eqref{Z32Schur}. In this case, the saddle point in the limit of $\alpha\to 0$ is $(x_1,y_1)=(0,0)$, i.e. the integration contour passes through the point (0,0), which is in the perfect agreement with the claim of sec.3.2.

The contribution associated with the second saddle point gives rise to a similarly looking generation function
        \begin{equation}\label{F2}
            F^{(x_2,y_2)}(p_1) = e^{f(x_2,y_2)} \sum_{n=0} (-1)^n\alpha^{6n} \frac{(6n+1)!! }{(6n)!!!(6n+1)}\left(1-4p_1\alpha^2 \right)^{-\frac{6n+1}{4}}
        \end{equation}
The key difference is in the exponential $e^f$ in front of the sum. While
\begin{equation}
            f(x_1,y_1)=\frac{p_1^3}{3}+\frac{1}{2} \alpha ^2 p_1^4+\alpha ^4 p_1^5+{7\over 3}\alpha^6p_1^6+\ldots
        \end{equation}
is regular at $ \alpha \to 0$, and starts with the third power of $p_1$ so that the exponential $e^{f(x_1,y_1)}$ vanishes in the limit of $\alpha \to 0$,
        \begin{equation}
            f(x_2,y_2)=-\frac{1}{6 \alpha ^6}+\frac{p_1}{\alpha ^4}-\frac{p_1^2}{\alpha ^2}-\frac{p_1^3}{3}-\frac{1}{2} \alpha ^2 p_1^4
            -\alpha ^4 p_1^5-{7\over 3}\alpha^6p_1^6+\ldots
        \end{equation}
contains all powers of $p_1$, and is singular in the limit of $\alpha\to 0$.

This means that this contribution of the second saddle point {\bf is suppressed in the limit of $\alpha\to 0$}, and is in accordance with the claim that the integral \eqref{WLZZfullInteralRep} is equal to zero when the integration contour does not pass through $(0,0)$.

Now one can sum up the two contributions, (\ref{sum}):
\be
F(p_1)= \beta_1 e^{f(x_1,y_1)}\cdot\sum_{n=0} \alpha^{6n} \frac{(6n+1)!! }{(6n)!!!(6n+1)}\left(1-4p_1\alpha^2 \right)^{-\frac{6n+1}{4}}+\nn\\
+\beta_2 e^{f(x_2,y_2)}\cdot\sum_{n=0}(-1)^n \alpha^{6n} \frac{(6n+1)!! }{(6n)!!!(6n+1)}\left(1-4p_1\alpha^2 \right)^{-\frac{6n+1}{4}}
\ee
and obtain
\be\label{bbeta}
b_{[r]}^{int}=\left.\frac{\partial^r}{\partial p_1^r}F(p_1)\right|_{p_1=0}
\ee
This answer should be compared with the coefficients (\ref{bRWLZZ32mi}): in the both cases, all $b_{[r]}$ are parameterized by two arbitrary parameters, $b_\emptyset$ and $b_{[1]}$ in (\ref{bRWLZZ32mi}), and $\beta_{1,2}$ in (\ref{bbeta}). One can check that upon identification
$b_\emptyset=b_\emptyset^{int}$ and $b_{[1]}=b_{[1]}^{int}$ in these two cases, all other $b_{[r]}$'s also coincide: $b_{[r]}=b^{int}_{[r]}$. The identification in terms of integration contour parameters $\beta_{1,2}$ is given manifestly by
\be
b_\emptyset^{int}&=&\beta_1\sum_{n=0}\alpha^{6n}{(6n+1)!!\over (6n)!!!(6n+1)}+
\beta_2e^{-{1\over 6\alpha^2}}\cdot\sum_{n=0}(-1)^n\alpha^{6n}{(6n+1)!!\over (6n)!!!(6n+1)}\\
b_{[1]}^{int}&=&\beta_1\sum_{n=0}\alpha^{6n+2}{(6n+1)!!\over (6n)!!!}+\beta_2e^{-{1\over 6\alpha^2}}\cdot
\sum_{n=0}(-1)^n\alpha^{6n+2}{(6n+1)!!\over (6n)!!!}+
\beta_2e^{-{1\over 6\alpha^2}}\cdot\sum_{n=0}(-1)^n\alpha^{6n-4}{(6n+1)!!\over (6n)!!!(6n+1)}\nn
\ee
Now, for instance, at $\beta_2=0$,
\be
b_{[2]}^{int}=\beta_1\sum_{n=0}\alpha^{6n+4}{(6n+1)!!(6n+5)\over (6n)!!!}=
\frac{b_{[1]}^{int}}{\alpha ^4} - \frac{b_\emptyset^{int}}{\alpha ^2}
\ee
Similarly, at $\beta_1=0$
\be
b_{[2]}^{int}=5\beta_2e^{-{1\over 6\alpha^2}}\cdot\sum_{n=0}(-1)^n\alpha^{6n-8}{(6n+1)!!\over (6n)!!!(6n-1)(6n+1)(6n-5)}=\frac{b_{[1]}^{int}}{\alpha ^4} - \frac{b_\emptyset^{int}}{\alpha ^2}
\ee
Since the relation
\be\label{b2}
b_{[2]}^{int}=\frac{b_{[1]}^{int}}{\alpha ^4} - \frac{b_\emptyset^{int}}{\alpha ^2}
\ee
is linear, it is satisfied at arbitrary $\beta_{1,2}$.

One can similarly check that formulas (\ref{bRWLZZ32mi}) are fulfilled for other $b_{[r]}^{int}$ equally well.

Thus, we see that {\bf an arbitrary solution of this type (given by one-line Young diagrams at $N=1$) to the Ward identities is obtained by a suitable choice of integration contours} in the matrix model partition function (\ref{z23n1}).

\section{Conclusion}

In this paper, we studied the WLZZ models described by two-matrix models, and demonstrated that the space of solutions to the corresponding Ward identities with shifted variables can be described by matrix integrals only in the limit of $N\to\infty$, and then this space is parameterized by various possible choices of integration contours for $N$ eigenvalues. At the same time, at finite $N$, there are additional constraints for matrix integrals so that they describe only a subspace of the space of all solutions to the Ward identities. This picture is in complete agreement with earlier results \cite{AMM23}, where a similar program was realized for the Hermitian one-matrix model.

Here we studied in detail only the case when the dependence on one of the matrices of the two-matrix model is described by the cubic potential. Moreover, we considered the most detailed example of only the second shifted variable. However, this is not a problem to consider similarly other cases. Still, while in the one-matrix model case, there is a generic description of the space of solutions in terms of an arbitrary function, moreover, there is a manifest construction of operators acting on these space (so called check-operators \cite{AMM23,MMcho}), a similar construction for the WLZZ models yet has to be found.

Similarly unknown remains an extension of the Dijkgraaf-Vafa construction \cite{DV} to the WLZZ case. This latter is associated in the one-matrix case with the spectral curves and their quantization. A similar construction in the case of WLZZ models is much more involved \cite{MMsc} and should be also incorporated into the framework discussed in this paper similarly to \cite{MMcho}. We are planning to discuss all these important issues elsewhere.

\section*{Acknowledgements}

The work was
partially funded within the state assignment of the Institute for Information Transmission Problems of RAS.
Our work is also partly supported by the grant of the Foundation for the Advancement of Theoretical Physics
and Mathematics ``BASIS".

\end{document}